\documentclass[reprint,showpacs,preprintnumbers,pre,superscriptaddress]{revtex4-1}
\usepackage[T1]{fontenc}
\usepackage[latin9]{inputenc}
\usepackage{amsmath}
\usepackage{amssymb}
\usepackage{graphicx}
\usepackage{esint}

\makeatletter
\usepackage{bm}\usepackage{amsfonts}
\usepackage{color}

\makeatother

\begin{document}
\title{Geometric Bounds on the Finite-Time Performance of Active Machines}
\author{Geng Li}
\affiliation{School of Systems Science, Beijing Normal University, Beijing 100875,
China}
\author{Z. C. Tu}
\email{tuzc@bnu.edu.cn}

\affiliation{School of Physics and Astronomy, Beijing Normal University, and Key
Laboratory of Multiscale Spin Physics (Beijing Normal University),
Ministry of Education, Beijing 100875, China}
\begin{abstract}
Optimizing energy conversion in active matter remains a central challenge
in nonequilibrium physics. Here, we develop a unified thermodynamic
framework that characterizes the finite-time performance of interacting
active machines. We show that cyclic work admits a geometric decomposition
into an antisymmetric thermodynamic curvature, governing work extraction,
and a symmetric metric, controlling dissipation. Minimal-dissipation
protocols follow geodesics in parameter space, while optimal work
extraction deviates from them due to a curvature-induced, Lorentz-like
effect. This geometric structure directly determines the finite-time
scaling of work and dissipation, enabling a mapping onto Onsager-type
quasi-linear current--force relations. We show that both the maximal
efficiency and the efficiency at maximum power are governed by an
asymmetry parameter and a figure of merit, establishing a formal correspondence
between active machines and thermoelectric devices with broken time-reversal
symmetry. Our results reveal a fundamental geometric origin of energy-conversion
performance and provide a general framework for optimizing active
machines.
\end{abstract}
\maketitle
\emph{Introduction.}--Active systems, comprising self-driven units
that continuously dissipate environmental fuel, serve as a versatile
platform for exploring nonequilibrium physics across biological and
synthetic scales \cite{Marchetti2013,Bechinger2016}. Examples range
from bacterial suspensions \cite{Wu2000} and cytoskeletal assemblies
\cite{Schaller2010} to artificial microswimmers \cite{Buttinoni2012}
and catalytic colloids \cite{Palacci2013}. Unlike passive systems,
the persistent motion of these entities breaks detailed balance and
time-reversal symmetry \cite{Bowick2022}, enabling them to generate
mechanical work and perform autonomous tasks far from equilibrium
\cite{Seifert2011}. Understanding the energetic performance and optimal
control of such systems has thus emerged as a cornerstone of nonequilibrium
statistical mechanics and soft-matter engineering.

Recent advances in stochastic thermodynamics have provided powerful
tools for analyzing energy conversion and control in small-scale systems
\cite{Seifert2012,GueryOdelin2023}. For passive systems near equilibrium,
linear response theory combined with thermodynamic geometry has revealed
universal principles: the dissipation associated with finite-time
driving scales inversely with the protocol duration, defining a Riemannian
metric where optimal protocols correspond to geodesics \cite{Salamon1983,Andresen1988,Crooks2007,Sivak2012,Chen2021,Li2022}.
While these concepts have been successfully applied to molecular machines
and colloidal heat engines \cite{Deffner2020,Chen2022}, extending
them to active matter remains a formidable challenge. The intrinsic
activity fundamentally alters the response properties and energetic
balance \cite{Fodor2016,Mandal2017,DalCengio2019}, leaving it unclear
whether universal scaling laws---analogous to those in passive systems---hold
in general \cite{Holubec2020,Fodor2021,Davis2024,Wang2025}, and how
self-propulsion dictates the ultimate limits of efficiency and power
\cite{Hill1974,Juelicher1997,Pietzonka2016,Szamel2020}.

\begin{figure}[!htp]
\includegraphics{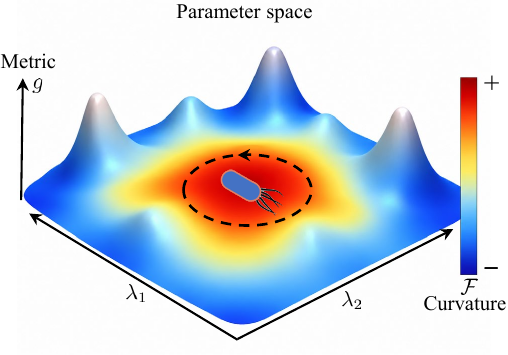} \caption{Schematic illustration of the geometric framework for active machines.
Cyclic driving in parameter space $(\lambda_{1},\lambda_{2})$ is
governed by two complementary geometric structures: an antisymmetric
thermodynamic curvature $\mathcal{F}$ responsible for geometric work
extraction and a symmetric thermodynamic metric $g$ governing irreversible
dissipation. Their competition determines the optimal finite-time
performance of active machines.}
\label{fig1}
\end{figure}

In this Letter, we develop a unified geometric framework for the finite-time
thermodynamic cycle of interacting active systems. As shown in Fig.$\ $\ref{fig1},
we show that cyclic work naturally decomposes into an antisymmetric
thermodynamic curvature responsible for geometric work extraction
and a symmetric dissipative kernel governing irreversible losses.
This geometric structure leads to universal finite-time scaling relations
for work and heat and establishes a direct connection between optimal
control and thermodynamic geometry. Interpreting the driven active
system as an energy-conversion machine, we derive Onsager-type quasi-linear
current--force relations and obtain universal bounds on both the
maximal efficiency and the efficiency at maximum power. These bounds
are governed by an asymmetry parameter and a figure of merit, and
take the same functional form as those of thermoelectric devices with
broken time-reversal symmetry. Unlike conventional thermoelectrics,
however, the figure of merit in active systems is not solely determined
by intrinsic material properties, but also by the geometry of the
driving protocol. This correspondence establishes a conceptual bridge
between active-matter thermodynamics and classical energy-conversion
theory, providing a geometric principle for optimizing the performance
of active machines.

\textit{Basic setup.}--We consider an active system composed of $N$
interacting overdamped self-propelled particles in three dimensions,
characterized by the microstate $\vec{\mathbf{r}}\equiv(\mathbf{r}_{1},\mathbf{r}_{2},\cdots,\mathbf{r}_{N})$.
The dynamics of particle $i$ obey
\begin{equation}
\gamma\dot{\mathbf{r}}_{i}=-\nabla_{i}U+\gamma\mathbf{v}_{i}+\boldsymbol{\xi}_{i},\label{eq:withorent}
\end{equation}
where the potential $U(\vec{\mathbf{r}},\boldsymbol{\lambda})$ includes
both externally controlled fields \cite{Snezhko2011,Palacci2013}
and interparticle interactions \cite{Theurkauff2012,Yan2016}, parameterized
by controllable work variables $\boldsymbol{\lambda}(t)\equiv(\lambda_{1},\lambda_{2},\cdots,\lambda_{M})$.
The self-propulsion velocity is specified by $\mathbf{v}_{i}\equiv v\mathbf{e}_{i}$,
with constant magnitude $v$. The orientation vector $\mathbf{e}_{i}$
evolves on the unit sphere according to $\ensuremath{\dot{\mathbf{e}}_{i}}=\ensuremath{\boldsymbol{\eta}_{i}}\ensuremath{\times}\ensuremath{\mathbf{e}_{i}}$,
leading to an exponential time correlation with persistence time $\tau_{p}\equiv\gamma_{r}/(2T)$.
The stochastic terms $\boldsymbol{\xi}_{i}$ and $\ensuremath{\boldsymbol{\eta}_{i}}$
are independent Gaussian white noises with zero mean and correlations
$\langle\boldsymbol{\xi}_{i}(t)\boldsymbol{\xi}_{j}(t')\rangle=2\gamma T\mathbf{1}\delta_{ij}\delta(t-t')$
and $\langle\boldsymbol{\eta}_{i}(t)\boldsymbol{\eta}_{j}(t')\rangle=2(T/\gamma_{r})\mathbf{1}\delta_{ij}\delta(t-t')$.
Here $T$ is bath temperature, and $\gamma$ and $\gamma_{r}$ are
the translational and rotational friction coefficients, respectively.
Throughout this work, the Boltzmann constant is set to unity.

Viewing the active system as a thermodynamic machine that converts
environmental fuel (e.g., ATP in biological systems) into useful work,
we consider cyclic driving over a finite interval $[0,\tau]$ with
periodic protocol $\boldsymbol{\lambda}(t)=\boldsymbol{\lambda}(t+\tau)$.
Following standard definitions of stochastic thermodynamics \cite{Jarzynski1997,Sekimoto1997},
the mean output work performed through the external control is $W\equiv-\int_{0}^{\tau}\langle\partial U/\partial\boldsymbol{\lambda}\rangle\cdot\dot{\boldsymbol{\lambda}}dt$
and the mean heat exchanged with the environment is $Q\equiv\sum_{i}\int_{0}^{\tau}\langle\dot{\mathbf{r}}_{i}\circ(-\gamma\dot{\mathbf{r}}_{i}+\boldsymbol{\xi}_{i})\rangle dt$
where $\langle\cdot\rangle$ denotes an ensemble average over stochastic
trajectories initialized from the steady state, and $\circ$ indicates
the Stratonovich convention for the product.

\textit{Linear response scaling.}--In the slow-driving regime, where
the control varies on time scales longer than intrinsic relaxation,
linear response theory allows us to approximate the expectation value
of a dynamical observable as \cite{Sivak2012,Davis2024}
\begin{equation}
\langle A(t)\rangle\approx\langle A\rangle_{s}+\boldsymbol{I}(A)\cdot\dot{\boldsymbol{\lambda}}(t),\label{eq:slowdriving}
\end{equation}
where $\boldsymbol{I}(A)\equiv\int_{0}^{\infty}\boldsymbol{\chi}(A;t',0)t'dt'$
with $\boldsymbol{\chi}(A;t',0)\equiv\delta\langle A(t')\rangle/\delta\boldsymbol{\lambda}(0)$
representing the response function of the observable $\langle A(t')\rangle$
to a perturbation in the control parameters $\boldsymbol{\lambda}(0)$.
The detailed derivations of Eq.$\ $(\ref{eq:slowdriving}) are presented
in Supplemental Material \cite{Supple}.

Applying this expansion to the observables $\langle\partial U/\partial\boldsymbol{\lambda}\rangle$,
$\langle U\rangle$, and $\langle f\rangle$ gives
\begin{equation}
W\approx-\int_{0}^{\tau}\langle\partial U/\partial\boldsymbol{\lambda}\rangle_{s}\cdot\dot{\boldsymbol{\lambda}}dt-\int_{0}^{\tau}\dot{\boldsymbol{\lambda}}^{T}\cdot\boldsymbol{I}(\partial U/\partial\boldsymbol{\lambda})\cdot\dot{\boldsymbol{\lambda}}dt\label{eq:linearwork}
\end{equation}
and
\begin{align}
Q= & \Delta\langle U\rangle-\int_{0}^{\tau}(\langle\partial U/\partial\boldsymbol{\lambda}\rangle\cdot\dot{\boldsymbol{\lambda}}+\langle f\rangle)dt\nonumber \\
\approx & \boldsymbol{I}(U)\cdot\dot{\boldsymbol{\lambda}}(\tau)-\int_{0}^{\tau}\dot{\boldsymbol{\lambda}}^{T}\cdot\langle\partial U/\partial\boldsymbol{\lambda}\rangle_{s}dt-\int_{0}^{\tau}\boldsymbol{I}(f)\cdot\dot{\boldsymbol{\lambda}}dt\nonumber \\
 & -\int_{0}^{\tau}\dot{\boldsymbol{\lambda}}^{T}\cdot\boldsymbol{I}(\partial U/\partial\boldsymbol{\lambda})\cdot\dot{\boldsymbol{\lambda}}dt-\int_{0}^{\tau}\langle f\rangle_{s}dt,\label{eq:linearheat}
\end{align}
where $\Delta\langle U\rangle\equiv\langle U(\tau)\rangle-\langle U(0)\rangle_{s}$
denotes the mean potential difference with $\langle\cdot\rangle_{s}$
being steady-state averaging and $f\equiv\gamma\sum_{i}\dot{\mathbf{r}}_{i}\circ\mathbf{v}_{i}$
represents the rate of energy injection from the active propulsion.
For the commonly adopted protocol design $\boldsymbol{\lambda}(t)=\boldsymbol{\lambda}(t/\tau)$
\cite{Campo2013}, it is convenient to introduce the rescaled time
variable $u=t/\tau$. Defining $\boldsymbol{\lambda}'\equiv d\boldsymbol{\lambda}/du$,
the work and heat acquire the finite-time thermodynamic scaling forms
\begin{equation}
W=\Gamma-\frac{\Sigma}{\tau},\;\;\;Q\approx\Gamma-\Psi-\frac{\Sigma}{\tau}-\Omega\tau.\label{eq:rescalwork}
\end{equation}
Here $\Gamma\equiv-\int_{0}^{1}\langle\partial U/\partial\boldsymbol{\lambda}\rangle_{s}\cdot[\boldsymbol{\lambda}']du$
is the quasistatic output work, $\Sigma\equiv\int_{0}^{1}[\boldsymbol{\lambda}']^{T}\cdot\boldsymbol{I}(\partial U/\partial\boldsymbol{\lambda})\cdot[\boldsymbol{\lambda}']du$
denotes the rescaled dissipation, $\Psi\equiv\int_{0}^{1}\boldsymbol{I}(f)\cdot[\boldsymbol{\lambda}']du$
quantifies the excess active energy input induced by the control protocol,
and $\Omega\equiv\int_{0}^{1}\langle f\rangle_{s}du$ characterizes
steady energy injection from activity. The boundary dissipation contribution
$\boldsymbol{I}(U)\cdot[\boldsymbol{\lambda}'](1)$ has been neglected,
since it is subleading compared with the integrated dissipation term
$\Sigma$ in the large-$\tau$ limit. Importantly, both the steady-state
contribution $\langle A\rangle_{s}$ and the response kernel $\boldsymbol{I}(A)$
depend solely on the intrinsic active dynamics and is independent
of the driving rate $\dot{\boldsymbol{\lambda}}$. 

Notably, the cyclic work contribution admits a natural geometric decomposition.
For a closed control protocol $\mathcal{C}$, the quasistatic work
can be expressed as $\Gamma=-\sum_{\mu}\oint_{\mathcal{C}}\langle\partial_{\mu}U\rangle_{s}d\lambda_{\mu}=\sum_{\mu\nu}\iint\mathcal{F}_{\mu\nu}d\lambda_{\mu}\wedge d\lambda_{\nu}$
where the antisymmetric thermodynamic curvature is defined as $\mathcal{F}_{\mu\nu}\equiv\partial_{\nu}\langle\partial_{\mu}U\rangle_{s}-\partial_{\mu}\langle\partial_{\nu}U\rangle_{s}$
with $\partial_{\mu}\equiv\partial/\partial\lambda_{\mu}$. This curvature
quantifies the geometric flux enclosed by the control cycle $\mathcal{C}$,
serving as a thermodynamic analog to the Berry curvature \cite{Berry1984,Wang2024,Fei2026}.
In equilibrium systems satisfying detailed balance, the quasistatic
generalized force derives from a scalar potential, implying $\mathcal{F}_{\mu\nu}=0$.
Consequently, no net work can be extracted through cyclic quasistatic
driving. By contrast, active nonequilibrium systems generally exhibit
nonvanishing thermodynamic curvature due to broken time-reversal symmetry,
thereby enabling geometric work extraction. 

In contrast to the antisymmetric curvature contribution, the finite-time
dissipation is governed by the symmetric part of the response tensor,
$g_{\mu\nu}=(I_{\mu\nu}+I_{\nu\mu})/2$ which defines a symmetric
quadratic form in the control-parameter space, $\Sigma=\int_{0}^{1}g_{\mu\nu}\lambda_{\mu}'\lambda_{\nu}'$.
For equilibrium systems satisfying detailed balance, $g_{\mu\nu}$
is positive definite and therefore induces a genuine Riemannian metric
structure, such that minimum-dissipation protocols correspond to geodesics
in parameter space \cite{Salamon1983,Andresen1988,Crooks2007,Sivak2012,Chen2021,Li2022}.
However, active nonequilibrium systems generally violate detailed
balance, and the symmetric response tensor need not remain positive
definite. In this case, $g_{\mu\nu}$ should be interpreted more generally
as an effective dissipation kernel rather than a strict thermodynamic
metric. Nevertheless, the competition between the symmetric dissipative
tensor and the antisymmetric thermodynamic curvature continues to
determine the optimal driving cycle. The curvature acts analogously
to a Lorentz-like force in parameter space, bending trajectories away
from purely dissipative geodesics and enabling geometric work extraction
unique to active systems \cite{Supple}.

The scaling relations of $Q$ in Eq.$\ $(\ref{eq:rescalwork}) are
consistent with the general expectations for optimal control of active
matter \cite{Davis2024}, which rely on a second-order weak-driving
expansion within linear response theory. Here we show that the same
scaling relations also emerge for general protocol designs of the
form $\boldsymbol{\lambda}(t)=\boldsymbol{\lambda}(t/\tau)$ while
keeping only the leading-order slow-driving contribution.

\textit{Quasi-linear current-force relations.}--Treating the active
system as a thermodynamic machine, we can formulate generalized current-force
relations that characterize its response to external driving. Positive
work extraction requires $\Gamma>\Sigma/\tau$. Based on the mechanism
of the active machine, the output power $P\equiv W/\tau$ associated
with external control can be expressed as the product of a generalized
mechanical current $J_{m}$ and force $X_{m}$: 
\begin{equation}
P=J_{m}X_{m}=\frac{\Gamma}{\tau}-\frac{\Sigma}{\tau^{2}}.\label{eq:outpower}
\end{equation}
Following finite-time thermodynamics \cite{Izumida2010,Sheng2014,Tu2020},
the cycle frequency $1/\tau$ is identified as a generalized mechanical
current $J_{m}$ while the conjugate mechanical force is defined as
$X_{m}\equiv W$. The active machine is powered by internal activity,
which generates an energy input rate $(W+T\Delta S_{\mathrm{tot}})/\tau$,
where the entropy production of the working process is given by $\Delta S_{\mathrm{tot}}\equiv\Delta S-Q/T$.
In the linear-response regime, the system entropy difference $\Delta S\approx0$,
and the active contribution can also be cast in a current--force
form,
\begin{equation}
J_{a}X_{a}=\frac{\Psi}{\tau}+\Omega,\label{eq:activeerate}
\end{equation}
where we define the active force $X_{a}\equiv\Gamma$, with conjugate
current $J_{a}$. Combining the above expressions yields Onsager-type
quasi-linear relations,
\begin{align}
\left(\begin{array}{c}
J_{m}\\
J_{a}
\end{array}\right) & =\left(\begin{array}{cc}
L_{mm} & L_{ma}\\
L_{am} & L_{aa}
\end{array}\right)\left(\begin{array}{c}
X_{m}\\
X_{a}
\end{array}\right),\label{eq:currentforc}
\end{align}
with generalized transport coefficients 
\begin{align}
L_{mm} & =L_{ma}=\frac{1}{\Sigma},\text{\ \ensuremath{\ \text{\ }}}L_{am}=\frac{1}{r}\frac{1}{\Sigma},\nonumber \\
L_{aa} & =\frac{1}{r}\frac{1}{\Sigma}+\frac{\Omega}{\Gamma^{2}},\label{eq:onsagercoef}
\end{align}
where $r\equiv L_{ma}/L_{am}=\Gamma/\Psi$ is the asymmetry parameter
quantifying the degree of reciprocity breaking in the quasi-linear
response relations. The condition $r\ne1$ reflects the breaking of
time-reversal symmetry in active systems, originating from the persistent
motion of self-propelled particles \cite{Bowick2022}. Unlike passive
systems constrained by detailed balance, this intrinsic persistence
leads to a violation of Onsager reciprocity at the macroscopic level.
This establishes a direct link between microscopic irreversibility
and macroscopic transport asymmetry, which ultimately constrains the
performance of active machines.

In contrast to linear nonequilibrium thermodynamics, where Onsager
coefficients are force-independent \cite{Groot2013}, the generalized
coefficients here depend explicitly on the active force $X_{a}$,
reflecting the inherently nonequilibrium nature of the system. Physically,
this dependence encodes energy leakage during operation, implying
that internal activity cannot be fully converted into useful work.
In the present setup, where the self-propulsion speed $v$ is fixed,
$X_{a}$ becomes constant with given protocol shape $\mathcal{C}$,
and the quasi-linear relations can be treated as effectively linear. 

\begin{figure*}[!tp]
\includegraphics{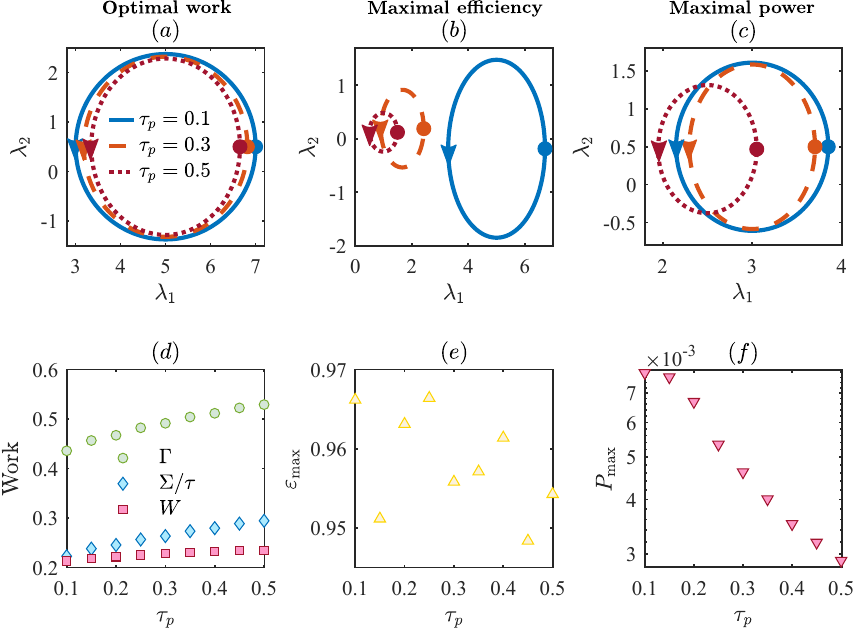} \caption{Geometric optimization of active-machine performance. Parameters are
fixed at $\gamma=1$, $T=1$, and $v=1$ throughout. Panels (a)-(c)
show the optimal cycles in the control space $(\lambda_{1},\lambda_{2})$
that maximize the extracted work, efficiency, and power, respectively,
for persistence times $\tau_{p}=0.1,$ $0.3$, and $0.5$. Symbols
mark the cycle starting points and arrows indicate the driving directions.
Increasing persistence shifts the balance between thermodynamic curvature,
which promotes geometric work extraction, and the dissipative kernel,
which penalizes finite-time driving, causing the optimal cycles to
shrink toward low-dissipation regions. Panels (d)-(f) show the resulting
optimal work, maximal efficiency, and maximal power. While stronger
persistence enhances the achievable geometric work, the maximal efficiency
varies only weakly and the maximal power decreases. These results
demonstrate that the finite-time performance of active machines is
controlled by the competition between geometric pumping and dissipation.}
\label{fig2}
\end{figure*}

\textit{Optimal performance of active machines.}--A central goal
in the study of energy-conversion systems is to determine the maximal
efficiency and power of a machine, as well as the conditions under
which these performance limits are achieved. For an active machine,
the efficiency is defined as the ratio between the output work and
the total energetic cost \cite{Hill1974,Juelicher1997,Pietzonka2016,Szamel2020,Fodor2021,Wang2025}:
\begin{equation}
\varepsilon\equiv\frac{J_{m}X_{m}}{J_{a}X_{a}}=\frac{\Gamma-\Sigma/\tau}{\Psi+\Omega\tau}.\label{eq:activeffe}
\end{equation}
Optimizing the efficiency $\varepsilon=\varepsilon(\mathcal{C},\tau)$
involves two independent controls: the protocol shape $\mathcal{C}$
in parameter space and the cycle duration $\tau$, naturally leading
to a hierarchical optimization. For a fixed protocol shape $\mathcal{C}$,
maximizing the efficiency with respect to $\tau$ yields
\begin{equation}
\varepsilon_{\mathrm{max}}=r\frac{\sqrt{\text{\ensuremath{\phi}}+1}-1}{\sqrt{\text{\ensuremath{\phi}}+1}+1},\label{eq:maxeff}
\end{equation}
where 
\begin{equation}
\phi\equiv\frac{L_{ma}L_{am}}{L_{mm}L_{aa}-L_{ma}L_{am}}=\frac{\Gamma\Psi}{\Sigma\Omega},\label{eq:figuremerit}
\end{equation}
plays a role analogous to the thermoelectric figure of merit $ZT$
\cite{Mahan1996,Broeck2005,Snyder2008,Esposito2009,Shakouri2011,Benenti2011,Tritt2011,Jiang2014,Jiang2015},
and the functional form of $\varepsilon_{\mathrm{max}}$ is identical
to that found in thermoelectric systems with broken time-reversal
symmetry. A key distinction, however, is that neither $r$ nor $\phi$
is an intrinsic material parameter. Both depend explicitly on the
protocol shape $\mathcal{C}$ through the geometric work $\Gamma$,
the active energetic cost $\Psi$ and $\Omega$, and the dissipative
contribution $\Sigma$. Consequently, the maximal efficiency is not
fixed by the properties of the active medium alone but can be systematically
enhanced through geometric optimization of the control cycle. Because
both $r$ nor $\phi$ vary with $\mathcal{C}$, the efficiency optimization
reflects a nontrivial competition between geometric work extraction,
active energetic expenditure, and finite-time dissipation. The optimal
protocol is therefore obtained numerically by directly maximizing
$\varepsilon_{\mathrm{max}}(\mathcal{C})$, as detailed in Supplemental
Material \cite{Supple}. These results reveal that geometry serves
not merely as a descriptor of active-machine dynamics, but as a control
resource that determines and enhances their achievable efficiency.

Another important indicator of machine performance is the efficiency
at maximum power (EMP). The output power $P=P(\mathcal{C},\tau)$,
defined in Eq.$\ $(\ref{eq:outpower}), can likewise be optimized
hierarchically. For a fixed protocol $\mathcal{C}$ with positive
dissipation $\Sigma>0$, maximizing the power with respect to $\tau$
gives
\begin{equation}
P_{\mathrm{max}}=\frac{L_{ma}^{2}}{4L_{mm}}X_{a}^{2}=\frac{\Gamma^{2}}{4\Sigma},\label{eq:maxpower}
\end{equation}
which is determined entirely by the balance between geometric work
extraction and finite-time dissipation. The corresponding efficiency
at maximum power is 
\begin{equation}
\varepsilon_{\mathrm{EMP}}=\frac{r\phi}{2(2+\phi)}.\label{eq:effmaxpow}
\end{equation}

These expressions retain the same mathematical structure as those
of thermoelectric devices, but here the underlying quantities are
geometric in origin. In particular, the maximum power is governed
by the competition between the thermodynamic curvature, which generates
geometric work through $\Gamma$, and the thermodynamic metric, which
penalizes finite-time driving through $\Sigma$. Consequently, optimizing
power amounts to identifying protocol shapes that achieve the optimal
balance between these two geometric contributions. This geometric
competition ultimately controls the finite-time performance of active
machines and determines the attainable trade-off between power output
and efficiency.

\emph{Example.}--To illustrate the general framework, we consider
a two-dimensional active Brownian particle $\mathbf{r}=(x,y)$ confined
by a harmonic potential $U=\lambda_{1}(x^{2}+y^{2})/2+\lambda_{2}xy$,
where the externally controlled parameters $\boldsymbol{\lambda}(t)\equiv(\lambda_{1},\lambda_{2})$
govern the isotropic confinement strength and the coupling between
the two spatial directions, respectively. The overdamped dynamics
obey 
\begin{align}
\gamma\dot{x} & =-\lambda_{1}x-\lambda_{2}y+\gamma v\cos\theta+\xi_{x}(t),\nonumber \\
\gamma\dot{y} & =-\lambda_{2}x-\lambda_{1}y+\gamma v\sin\theta+\xi_{y}(t),\label{eq:exampedynamics}
\end{align}
with rotational diffusion $\dot{\theta}=\eta(t)$. In equilibrium
$(v=0)$, the generalized force derives from the scalar free energy,
leading to a vanishing thermodynamic curvature, $\mathcal{F}_{12}=0$,
such that no quasistatic work can be extracted from cyclic driving.
By contrast, activity $(v\ne0)$ breaks detailed balance and generates
a finite thermodynamic curvature, thereby enabling geometric work
extraction through cyclic modulation of the control parameters.

The harmonic structure allows the steady-state correlation and response
functions to be obtained analytically. Consequently, the geometric
work $\Gamma$, the rescaled dissipation $\Sigma$, and the active
energetic contributions $\Psi$ and $\Omega$ can all be evaluated
explicitly, with detailed expressions provided in Supplemental Material
\cite{Supple}. This enables a direct investigation of the competition
between geometric pumping and irreversible dissipation, and how this
competition determines the optimal performance of active machines.
To optimize the driving protocols, we parameterize the control cycle
using Fourier modes, $\lambda_{\mu}(u)=a_{\mu}^{(0)}+\sum_{n=1}^{N_{F}}[a_{\mu}^{(n)}\cos(2\pi nu)+b_{\mu}^{(n)}\sin(2\pi nu)]$,
with rescaled time $u=t/\tau\in[0,1]$. The Fourier coefficients are
optimized subject to stability constraints $\lambda_{1}>|\lambda_{2}|$,
ensuring confinement throughout the cycle. Figures$\ $\ref{fig2}(a)-(c)
show the optimal protocols maximizing the output work, efficiency,
and power, respectively, for persistence times $\tau_{p}=0.1,$ $0.3$,
and $0.5$. As $\tau_{p}$ increases, the optimal cycles progressively
contract toward regions of weaker dissipation. This behavior reflects
the increasing importance of the dissipative kernel relative to the
geometric work contribution, causing the optimal protocols to favor
low-dissipation regions of parameter space. Nevertheless, the enhanced
persistence simultaneously strengthens the nonequilibrium geometric
effects responsible for work extraction.

The corresponding optimal work, maximal efficiency, and maximal power
are shown in Figs.$\ $\ref{fig2}(d)-(f). As the persistence time
increases, the optimal work increases monotonically, indicating enhanced
geometric pumping. By contrast, the maximal efficiency remains nearly
unchanged over the explored range of $\tau_{p}$, while the maximal
power gradually decreases due to the growing energetic cost associated
with finite-time driving. Throughout the simulations, we fix $\gamma=1$,
$T=1$, $v=1$, and $N_{F}=1$. For the work optimization shown in
Figs.$\ $(a) and$\ $(d), the protocol duration is fixed at $\tau=20$.

\emph{Conclusions.}--We have developed a unified geometric framework
for the finite-time thermodynamic cycle of active matter, establishing
a general description of nonequilibrium work extraction in terms of
thermodynamic curvature and dissipative geometric structures. Within
this framework, cyclic work naturally decomposes into an antisymmetric
curvature contribution responsible for geometric pumping and a symmetric
dissipative kernel governing irreversible losses. This geometric structure
determines the finite-time scaling of work and heat and provides a
direct connection between optimal control and nonequilibrium thermodynamics
in active systems. By mapping driven active systems onto Onsager-type
quasi-linear current--force relations, we derived universal bounds
on both the maximal efficiency and the efficiency at maximum power.
Remarkably, these bounds take the same functional form as those of
thermoelectric devices with broken time-reversal symmetry and are
governed by a dimensionless figure of merit $\phi$. Unlike the thermoelectric
figure of merit $ZT$, which is fixed by microscopic transport coefficients,
$\phi$ emerges from the interplay between nonequilibrium activity
and the geometry of the driving protocol. Consequently, the performance
of active machines is not solely constrained by intrinsic material
properties, but can be systematically optimized through geometric
control in parameter space. The decomposition of performance into
curvature-driven and metric-driven contributions arises naturally
from linear-response theory, suggesting that the geometric framework
developed here may provide a unified description of optimal control
and performance in a broad class of nonequilibrium steady-state systems.

\emph{Acknowledgement.}--This work is supported by the National Natural
Science Foundation of China (NSFC) (Grants No. 12405031 and No. 12475032).
G. L. acknowledges support from the Zhongying Young Scholars Program
of Beijing Normal University.

\bibliographystyle{apsrev4-1}
\bibliography{ref}

\end{document}


\title{Supplementary Material: Geometric Bounds on the Finite-Time Performance
of Active Machines}
\author{Geng Li}
\affiliation{School of Systems Science, Beijing Normal University, Beijing 100875,
China}
\author{Z. C. Tu}
\email{tuzc@bnu.edu.cn}

\affiliation{School of Physics and Astronomy, Beijing Normal University, and Key
Laboratory of Multiscale Spin Physics (Beijing Normal University),
Ministry of Education, Beijing 100875, China}

\maketitle
The supplementary materials are devoted to provide detailed derivations
in the main context.

\tableofcontents{}

\section{Linear response approximation under slow and weak driving protocols\label{SSec-one}}

In this section, we present detailed expressions of the linear response
function under slow and weak driving.

\subsection{Linear response approximation\label{SSec-oneA}}

Under slowly varying control protocols, linear response theory allows
us to express the expectation value of a dynamical observable $\langle A(t)\rangle$
in terms of response functions as
\begin{align}
\langle A(t)\rangle & \approx\langle A\rangle_{s}+\int_{-\infty}^{t}\boldsymbol{\chi}(A;t,t')\cdot[\boldsymbol{\lambda}(t)-\boldsymbol{\lambda}(t')]dt'\label{seq:linearresponse}
\end{align}
Expanding the control parameters around time $t'$,
\begin{equation}
\boldsymbol{\lambda}(t)\approx\boldsymbol{\lambda}(t')+\dot{\boldsymbol{\lambda}}(t')(t-t')+O((t-t')^{2}),\label{seq:protocolslow}
\end{equation}
and retaining terms up to first order, we obtain
\begin{equation}
\langle A(t)\rangle\approx\langle A\rangle_{s}+\int_{-\infty}^{t}\boldsymbol{\chi}(A;t,t')\cdot\dot{\boldsymbol{\lambda}}(t')(t-t')dt'.\label{seq:linearapp}
\end{equation}
Here, the response function is defined as $\boldsymbol{\chi}(A;t,t')\equiv\delta\langle A(t)\rangle/\delta\boldsymbol{\lambda}(t')$.
Introducing the variable change $t''=t-t'$, and expanding the control
velocity as
\begin{equation}
\dot{\boldsymbol{\lambda}}(t')=\dot{\boldsymbol{\lambda}}(t-t'')\approx\dot{\boldsymbol{\lambda}}(t)+O(\ddot{\boldsymbol{\lambda}}),\label{seq:protocolveco}
\end{equation}
we obtain a time-local approximation:
\begin{align}
\langle A(t)\rangle & \approx\langle A\rangle_{s}+\dot{\boldsymbol{\lambda}}(t)\cdot\int_{-\infty}^{t}\boldsymbol{\chi}(A;t,t')(t-t')dt'\nonumber \\
 & =\langle A\rangle_{s}+\dot{\boldsymbol{\lambda}}(t)\cdot\int_{0}^{\infty}\boldsymbol{\chi}(A;t'',0)t''dt''\nonumber \\
 & =\langle A\rangle_{s}+\boldsymbol{I}(A)\cdot\dot{\boldsymbol{\lambda}}(t),\label{seq:timelocallinear}
\end{align}
where $\boldsymbol{I}(A)\equiv\int_{0}^{\infty}\boldsymbol{\chi}(A;t'',0)t''dt''$
represents the response kernel. This expression is equivalent to Eq.$\ $(5)
in the main text.

\subsection{Response function\label{SSec-oneB}}

The expectation value of a dynamical observable $\langle A(t)\rangle$
can be expressed in the path-integral representation as
\begin{align}
\langle A(t)\rangle & =\Pi_{i}\iint D[\mathbf{v}_{i}]D[\boldsymbol{\xi}_{i}]A[t;\boldsymbol{\xi}_{i},\mathbf{v}_{i}]P_{v}[\mathbf{v}_{i}]P_{\xi}[\boldsymbol{\xi}_{i}],\label{seq:pathintegres}
\end{align}
where $P_{\xi}[\boldsymbol{\xi}_{i}]$ and $P_{v}[\mathbf{v}_{i}]$
represent the probability functionals of the thermal noise $\boldsymbol{\xi}_{i}$
and the active noise $\mathbf{v}_{i}$, respectively. Starting from
the Langevin dynamics in Eq.$\ $(1), we insert the identity
\begin{align}
1 & =\Pi_{i}\int D[\boldsymbol{\xi}_{i}]\delta(\dot{\mathbf{r}}_{i}+\gamma^{-1}\nabla_{i}U-\gamma^{-1}\boldsymbol{\xi}_{i}-\mathbf{v}_{i}),\label{seq:pathiden}
\end{align}
and transform the integration variables from $\boldsymbol{\xi}_{i}$
to $\mathbf{r}_{i}$, yielding

\begin{align}
1 & =\Pi_{i}\int D[\mathbf{r}_{i}]J[\mathbf{r}_{i}]\delta(\dot{\mathbf{r}}_{i}+\gamma^{-1}\nabla_{i}U-\gamma^{-1}\boldsymbol{\xi}_{i}-\mathbf{v}_{i}),\label{seq:yakebi}
\end{align}
where $J[\mathbf{r}_{i}]\equiv\mathrm{det}(\delta\boldsymbol{\xi}_{i}(t)/\delta\mathbf{r}_{i}(t'))$
is the corresponding Jacobian. Substituting this identity into the
path integral and integrating out the thermal noise leads to
\begin{align}
\langle A(t)\rangle & =\Pi_{i}\iiint D[\mathbf{v}_{i}]D[\boldsymbol{\xi}_{i}]D[\mathbf{r}_{i}]J[\mathbf{r}_{i}]A[t;\boldsymbol{\xi}_{i},\mathbf{v}_{i}]P_{v}[\mathbf{v}_{i}]P_{\xi}[\boldsymbol{\xi}_{i}]\delta(\dot{\mathbf{r}}_{i}+\gamma^{-1}\nabla_{i}U-\gamma^{-1}\boldsymbol{\xi}_{i}-\mathbf{v}_{i})\nonumber \\
 & =\mathcal{N}^{-1}\Pi_{i}\iint D[\mathbf{v}_{i}]D[\mathbf{r}_{i}]J[\mathbf{r}_{i}]P_{v}[\mathbf{v}_{i}]A[t;\mathbf{r}_{i},\mathbf{v}_{i}]e^{-\mathcal{S}[\mathbf{r}_{i},\mathbf{v}_{i}]},\label{seq:integthermalno}
\end{align}
 where $\mathcal{S}\equiv[\gamma/(4T)]\int dt(\dot{\mathbf{r}}_{i}+\gamma^{-1}\nabla_{i}U-\mathbf{v}_{i})^{2}$
denotes the Onsager-Machlup action \citep{Onsager1953}, and
\begin{equation}
\mathcal{N}\equiv\Pi_{i}\iint D[\mathbf{v}_{i}]D[\mathbf{r}_{i}]J[\mathbf{r}_{i}]P_{v}[\mathbf{v}_{i}]e^{-\mathcal{S}[\mathbf{r}_{i},\mathbf{v}_{i}]}\label{seq:normconst}
\end{equation}
 ensures normalization.

Differentiating the path-integral expression in Eq.$\ $(\ref{seq:integthermalno})
yields the response function as
\begin{align}
\boldsymbol{\chi}(A;t,t') & \equiv\frac{\delta\langle A(t)\rangle}{\delta\boldsymbol{\lambda}(t')}\nonumber \\
 & =\Pi_{i}\iint D[\mathbf{v}_{i}]D[\mathbf{r}_{i}]J[\mathbf{r}_{i}]P_{v}[\mathbf{v}_{i}]A[t;\mathbf{r}_{i},\mathbf{v}_{i}]\frac{\delta(\mathcal{N}^{-1}e^{-\mathcal{S}})}{\delta\boldsymbol{\lambda}(t')}\nonumber \\
 & =-\Pi_{i}\iint D[\mathbf{v}_{i}]D[\mathbf{r}_{i}]J[\mathbf{r}_{i}]P_{v}[\mathbf{v}_{i}]A[t;\mathbf{r}_{i},\mathbf{v}_{i}]\mathcal{N}^{-1}e^{-\mathcal{S}}(\frac{\delta\mathcal{S}}{\delta\boldsymbol{\lambda}(t')}-\langle\frac{\delta\mathcal{S}}{\delta\boldsymbol{\lambda}(t')}\rangle)\nonumber \\
 & =-\langle A(t)\frac{\delta\mathcal{S}}{\delta\boldsymbol{\lambda}(t')}\rangle+\langle A(t)\rangle\langle\frac{\delta\mathcal{S}}{\delta\boldsymbol{\lambda}(t')}\rangle,\label{seq:resfunS}
\end{align}
where we have used
\begin{align}
\frac{\delta\mathcal{N}}{\delta\boldsymbol{\lambda}(t')} & =-\Pi_{i}\iint D[\mathbf{v}_{i}]D[\mathbf{r}_{i}]J[\mathbf{r}_{i}]P_{v}[\mathbf{v}_{i}]e^{-\mathcal{S}}\frac{\delta\mathcal{S}}{\delta\boldsymbol{\lambda}(t')}\nonumber \\
 & =-\mathcal{N}\langle\frac{\delta\mathcal{S}}{\delta\boldsymbol{\lambda}(t')}\rangle.\label{seq:deffalphanoe}
\end{align}
For the unperturbed dynamics with constant control parameters, the
functional derivative of the action becomes
\begin{align}
\frac{\delta\mathcal{S}}{\delta\boldsymbol{\lambda}(t')} & =\frac{1}{2T}\nabla_{i}(\frac{\partial U}{\partial\boldsymbol{\lambda}})(\dot{\mathbf{r}}_{i}+\gamma^{-1}\nabla_{i}U-\mathbf{v}_{i})\nonumber \\
 & =\frac{1}{2T}[\nabla_{i}(\frac{\partial U}{\partial\boldsymbol{\lambda}})\cdot\dot{\mathbf{r}}_{i}+\nabla_{i}(\frac{\partial U}{\partial\boldsymbol{\lambda}})\cdot(\gamma^{-1}\nabla_{i}U-\mathbf{v}_{i})]\nonumber \\
 & =\frac{1}{2T}[\frac{d}{dt'}(\frac{\partial U}{\partial\boldsymbol{\lambda}})+\nabla_{i}(\frac{\partial U}{\partial\boldsymbol{\lambda}})\cdot(\gamma^{-1}\nabla_{i}U-\mathbf{v}_{i})].\label{seq:deffsrespon}
\end{align}
Substituting this result into the response function finally gives
\begin{align}
\boldsymbol{\chi}(A;t,t') & =\frac{1}{2T}[-\frac{d}{dt'}\langle A(t)(\frac{\partial U}{\partial\boldsymbol{\lambda}})_{t'}\rangle+\langle A\rangle\frac{d}{dt'}\langle(\frac{\partial U}{\partial\boldsymbol{\lambda}})_{t'}\rangle-\langle A(t)\nabla_{i}(\frac{\partial U}{\partial\boldsymbol{\lambda}})_{t'}\cdot(\gamma^{-1}\nabla_{i}U-\mathbf{v}_{i})_{t'}\rangle+\langle A\rangle\langle\nabla_{i}(\frac{\partial U}{\partial\boldsymbol{\lambda}})\cdot(\gamma^{-1}\nabla_{i}U-\mathbf{v}_{i})\rangle].\label{seq:finalresponse}
\end{align}
For a single control parameter, $\boldsymbol{\lambda}\to\lambda$,
this expression reduces to the response formula derived by Davis \textit{et
al. }\citep{Davis2024}. Equation$\ $(\ref{seq:finalresponse}) provides
a generalized fluctuation--response relation for active matter under
multidimensional control protocols and serves as the starting point
for the thermodynamic geometric formulation developed below. The response
kernel then follows as
\begin{align}
\boldsymbol{I}(A)\equiv & \int_{0}^{\infty}\boldsymbol{\chi}(A;t,0)tdt\nonumber \\
= & \frac{1}{2T}\int_{0}^{\infty}\{\langle A(t)(\frac{\partial U}{\partial\boldsymbol{\lambda}})_{t'=0}\rangle-\langle A(t)\rangle\langle(\frac{\partial U}{\partial\boldsymbol{\lambda}})_{t'=0}\rangle-t[\langle A(t)\nabla_{i}(\frac{\partial U}{\partial\boldsymbol{\lambda}})_{t'=0}\cdot(\gamma^{-1}\nabla_{i}U-\mathbf{v}_{i})_{t'=0}\rangle\nonumber \\
 & -\langle A\rangle\langle\nabla_{i}(\frac{\partial U}{\partial\boldsymbol{\lambda}})\cdot(\gamma^{-1}\nabla_{i}U-\mathbf{v}_{i})\rangle]\}.\label{seq:resknerl}
\end{align}

\section{Linear response approximation for the mean work and heat\label{SSec-two}}

According to stochastic thermodynamics \citep{Jarzynski1997,Sekimoto1997},
the mean output work is defined as
\begin{equation}
W\equiv-\int_{0}^{\tau}\langle\partial U/\partial\boldsymbol{\lambda}\rangle\cdot\dot{\boldsymbol{\lambda}}dt\label{seq:meanwork}
\end{equation}
and the mean heat absorbed from the environment is
\begin{equation}
Q\equiv\sum_{i}\int_{0}^{\tau}\langle\dot{\mathbf{r}}_{i}\circ(-\gamma\dot{\mathbf{r}}_{i}+\boldsymbol{\xi}_{i})\rangle dt.\label{seq:meanheat}
\end{equation}
Using the first-law relation, the heat can be rewritten as
\begin{equation}
Q=\Delta\langle U\rangle-\int_{0}^{\tau}(\langle\partial U/\partial\boldsymbol{\lambda}\rangle\cdot\dot{\boldsymbol{\lambda}}+\langle f\rangle)dt,\label{seq:firstlawre}
\end{equation}
where $\Delta\langle U\rangle\equiv\langle U(\tau)\rangle-\langle U(0)\rangle_{s}$
denotes the mean potential difference. Under the linear response approximation,
the relevant observables in Eqs.$\ $(\ref{seq:meanwork}) and$\ $(\ref{seq:meanheat})
can be expanded as
\begin{align}
\langle\partial U/\partial\boldsymbol{\lambda}\rangle & \approx\langle\partial U/\partial\boldsymbol{\lambda}\rangle_{s}+\boldsymbol{I}(\partial U/\partial\boldsymbol{\lambda})\cdot\dot{\boldsymbol{\lambda}}(t),\nonumber \\
\langle U(\tau)\rangle & \approx\langle U(\tau)\rangle_{s}+\boldsymbol{I}(U)\cdot\dot{\boldsymbol{\lambda}}(\tau),\nonumber \\
\langle f\rangle & \approx\langle f\rangle_{s}+\boldsymbol{I}(f)\cdot\dot{\boldsymbol{\lambda}}(t),\label{seq:linearresobser}
\end{align}
where $\boldsymbol{I}(\partial U/\partial\boldsymbol{\lambda})$,
$\boldsymbol{I}(U)$, and $\boldsymbol{I}(f)$ denote the corresponding
response kernels. Substituting these expressions into the definitions
of work and heat, we obtain
\begin{equation}
W\approx-\int_{0}^{\tau}\langle\partial U/\partial\boldsymbol{\lambda}\rangle_{s}\cdot\dot{\boldsymbol{\lambda}}dt-\int_{0}^{\tau}\dot{\boldsymbol{\lambda}}^{T}\cdot\boldsymbol{I}(\partial U/\partial\boldsymbol{\lambda})\cdot\dot{\boldsymbol{\lambda}}dt\label{seq:linearmeanwork}
\end{equation}
and
\begin{align}
Q\approx & \boldsymbol{I}(U)\cdot\dot{\boldsymbol{\lambda}}(\tau)-\int_{0}^{\tau}\langle\partial U/\partial\boldsymbol{\lambda}\rangle_{s}\cdot\dot{\boldsymbol{\lambda}}dt-\int_{0}^{\tau}\boldsymbol{I}(f)\cdot\dot{\boldsymbol{\lambda}}dt\nonumber \\
 & -\int_{0}^{\tau}\dot{\boldsymbol{\lambda}}^{T}\cdot\boldsymbol{I}(\partial U/\partial\boldsymbol{\lambda})\cdot\dot{\boldsymbol{\lambda}}dt-\int_{0}^{\tau}\langle f\rangle_{s}dt,\label{seq:linearmeanheat}
\end{align}
where $\Delta\langle U\rangle_{s}=0$ for cyclic processes in the
linear response approxiamtion. The results in Eqs.$\ $(\ref{seq:linearmeanwork})
and$\ $(\ref{seq:linearmeanheat}) directly correspond to Eqs$\ $(6)
and$\ $(7) in the main text.

\section{The scaling relations of the mean work and heat\label{SSec-three}}

In this section, we prove the scaling relations of the mean work and
heat within the linear response framework.

\subsection{Scaling of response kernels\label{SSec-threeA}}

In linear response theory, the response kernel associated with an
observable $A$ is defined as
\begin{align}
\boldsymbol{I}(A)\equiv\int_{0}^{\infty}\boldsymbol{\chi}(A;t'',0)t'dt', & \ \text{\ \ensuremath{\ }}\boldsymbol{\chi}(A;t,0)\equiv\delta\langle A(t)\rangle/\delta\boldsymbol{\lambda}(0).\label{seq:responsekernel}
\end{align}
By construction, the response function $\boldsymbol{\chi}$ characterizes
the system\textquoteright s response to an infinitesimal perturbation
around a reference state with fixed control parameters. As a result,
the kernel $\boldsymbol{I}(A)$ depends only on the intrinsic dynamical
properties of the system---such as the interaction potential, noise
statistics, and activity parameters---and is independent of the driving
rate $\dot{\boldsymbol{\lambda}}$.

For the commonly adopted protocol design $\boldsymbol{\lambda}(t)=\boldsymbol{\lambda}(t/\tau)$
\citep{Campo2013}, introducing the rescaled time $u=t/\tau$, the
control velocity scales as
\begin{equation}
\dot{\boldsymbol{\lambda}}(t)=\frac{1}{\tau}\boldsymbol{\lambda}'(u),\text{\ \ensuremath{\ \text{\ }}}\boldsymbol{\lambda}'\equiv d\boldsymbol{\lambda}/du.\label{seq:velocityscaling}
\end{equation}
Accordingly, the linear-response correction behaves as
\begin{equation}
\boldsymbol{I}(A)\cdot\dot{\boldsymbol{\lambda}}(t)=\frac{1}{\tau}\boldsymbol{I}(A)\cdot[\boldsymbol{\lambda}'](u),\label{seq:responsescaling}
\end{equation}
and its time integral satisfies
\begin{equation}
\int_{0}^{\tau}\boldsymbol{I}(A)\cdot\dot{\boldsymbol{\lambda}}dt=\int_{0}^{1}\boldsymbol{I}(A)\cdot[\boldsymbol{\lambda}']du,\label{seq:integralscaling}
\end{equation}
which is independent of the protocol duration $\tau$.

In contrast, the steady-state contribution $\langle A\rangle_{s}$,
evaluated at fixed control parameters, is independent of the driving
rate. Its time integral therefore scales as
\begin{equation}
\int_{0}^{\tau}\langle A\rangle_{s}dt=\tau\int_{0}^{1}\langle A\rangle_{s}du,\label{seq:tauplscaling}
\end{equation}
provided that $\langle A\rangle_{s}$ remains finite along the protocol.

Physically, this separation reflects two distinct contributions: the
steady-state term $\langle A\rangle_{s}$ describes persistent nonequilibrium
activity and gives rise to an extensive $O(\tau)$ contribution, while
the kernel $\boldsymbol{I}(A)$ plays a role analogous to a transport
coefficient, characterizing the intrinsic response of the active system
to external perturbations and yielding a finite $O(1)$ correction,
while the driving rate $\dot{\boldsymbol{\lambda}}$ acts as the conjugate
thermodynamic force.

\subsection{Universal scaling structure of work and heat\label{SSec-threeB}}

Based on the above analysis, the response relations in Eq.$\ $(\ref{seq:linearresobser})
take the form
\begin{align}
\langle\partial U/\partial\boldsymbol{\lambda}\rangle & \approx\langle\partial U/\partial\boldsymbol{\lambda}\rangle_{s}+\frac{1}{\tau}\boldsymbol{I}(\partial U/\partial\boldsymbol{\lambda})\cdot[\boldsymbol{\lambda}'](u),\nonumber \\
\langle U(\tau)\rangle & \approx\langle U(1)\rangle_{s}+\frac{1}{\tau}\boldsymbol{I}(U)\cdot[\boldsymbol{\lambda}'](1),\nonumber \\
\langle f\rangle & \approx\langle f\rangle_{s}+\frac{1}{\tau}\boldsymbol{I}(f)\cdot[\boldsymbol{\lambda}'](u).\label{seq:resuufscal}
\end{align}
Substituting these expressions into the approximate forms of work
and heat in Eqs.$\ $(\ref{seq:linearmeanwork}) and$\ $(\ref{seq:linearmeanheat}),
we obtain
\begin{equation}
W\approx-\int_{0}^{1}\langle\partial U/\partial\boldsymbol{\lambda}\rangle_{s}\cdot[\boldsymbol{\lambda}']du-\frac{1}{\tau}\int_{0}^{1}[\boldsymbol{\lambda}']^{T}\cdot\boldsymbol{I}(\partial U/\partial\boldsymbol{\lambda})\cdot[\boldsymbol{\lambda}']du\label{seq:scalingstructurework}
\end{equation}
and
\begin{align}
Q\approx & -\int_{0}^{1}\langle\partial U/\partial\boldsymbol{\lambda}\rangle_{s}\cdot[\boldsymbol{\lambda}']dt-\int_{0}^{1}\boldsymbol{I}(f)\cdot[\boldsymbol{\lambda}']du\nonumber \\
 & -\tau\int_{0}^{1}\langle f\rangle_{s}du+\frac{1}{\tau}\{\boldsymbol{I}(U)\cdot[\boldsymbol{\lambda}'](1)-\int_{0}^{1}[\boldsymbol{\lambda}']^{T}\cdot\boldsymbol{I}(\partial U/\partial\boldsymbol{\lambda})\cdot[\boldsymbol{\lambda}']du\}.\label{seq:scalingstructureheat}
\end{align}
These expressions reveal a universal scaling structure characterized
by three distinct contributions:
\begin{equation}
W=\Gamma-\frac{\Sigma}{\tau},\;\;\;Q\approx\Lambda-\frac{\Sigma}{\tau}-\Omega\tau,\label{seq:universcali}
\end{equation}
where the $\tau$-independent coefficients are
\begin{gather}
\Gamma\equiv-\int_{0}^{1}\langle\partial U/\partial\boldsymbol{\lambda}\rangle_{s}\cdot[\boldsymbol{\lambda}']du,\;\;\;\boldsymbol{\lambda}'\equiv d\boldsymbol{\lambda}/du,\nonumber \\
\Sigma\equiv\int_{0}^{1}[\boldsymbol{\lambda}']^{T}\cdot\boldsymbol{I}(\partial U/\partial\boldsymbol{\lambda})\cdot[\boldsymbol{\lambda}']du,\;\;\;\Omega\equiv\int_{0}^{1}\langle f\rangle_{s}du,\nonumber \\
\Lambda\equiv-\int_{0}^{1}\langle\partial U/\partial\boldsymbol{\lambda}\rangle_{s}\cdot[\boldsymbol{\lambda}']du-\int_{0}^{1}\boldsymbol{I}(f)\cdot[\boldsymbol{\lambda}']du.\label{eq:functiongasg}
\end{gather}
Here we have considered the fact that the boundary term $\boldsymbol{I}(U)\cdot[\boldsymbol{\lambda}'](1)$
is subleading compared with the integrated dissipation term and can
therefore be neglected for sufficiently large protocol duration $\tau$.
The leading $O(1/\tau)$ contribution then arises from the quadratic
form
\begin{equation}
\Sigma\equiv\int_{0}^{1}[\boldsymbol{\lambda}']^{T}\cdot\boldsymbol{I}(\partial U/\partial\boldsymbol{\lambda})\cdot[\boldsymbol{\lambda}']du,\label{seq:geometricdiss}
\end{equation}
which defines a geometric dissipation functional. The universal scaling
in Eq.$\ $(\ref{seq:universcali}) reduces to Eq.$\ $(8) in the
main text.

\section{Optimization over the performance of active machines\label{SSec-five}}

In this section, we derive the optimal protocols for maximizing the
output work, efficiency, and power of active machines.

\subsection{Optimization of output work\label{SSec-fiveA}}

According to the geometric decomposition
\begin{equation}
W(\mathcal{C},\tau)=\Gamma(\mathcal{C})-\frac{\Sigma(\mathcal{C})}{\tau},\label{seq:workdecomng}
\end{equation}
the maximal output work for a given cycle duration $\tau$ is achieved
by an optimal protocol $\mathcal{C}^{*}$. Performing a variational
optimization of the protocol shape yields the generalized geodesic
equation
\begin{equation}
\lambda_{\mu}^{''}+\sum_{\nu\kappa}\Gamma_{\nu\kappa}^{\mu}\lambda_{\nu}^{'}\lambda_{\kappa}^{'}=\frac{\tau}{2}\sum_{\nu}\mathcal{F}_{\nu}^{\mu}\lambda_{\nu}^{'},\label{seq:ggeuq}
\end{equation}
where $\Gamma_{\nu\kappa}^{\mu}\equiv\sum_{\iota}(g^{-1})_{\iota\mu}(\partial_{\kappa}g_{\iota\nu}+\partial_{\nu}g_{\iota\kappa}-\partial_{\iota}g_{\nu\kappa})/2$
is the Christoffel symbol associated with the thermodynamic metric
$g$, and $\mathcal{F}_{\nu}^{\mu}\equiv\sum_{\iota}(g^{-1})_{\text{\ensuremath{\iota\mu}}}\mathcal{F}_{\iota\nu}$
is the mixed-index thermodynamic curvature tensor. Owing to the antisymmetry
of $\mathcal{F}_{\iota\nu}$, the curvature-induced force is always
perpendicular to the protocol velocity in parameter space. It therefore
plays a role analogous to the Lorentz force in charged-particle dynamics,
deflecting the protocol away from dissipative geodesics while simultaneously
enabling geometric work extraction.

\subsection{Optimization of efficiency\label{SSec-fiveB}}

For a given protocol shape $\mathcal{C}$, the efficiency of the active
machine is
\begin{equation}
\varepsilon(\mathcal{C},\tau)=\frac{\Gamma(\mathcal{C})-\Sigma(\mathcal{C})/\tau}{\Psi(\mathcal{C})+\Omega(\mathcal{C})\tau},\label{seq:proparaeff}
\end{equation}
where $\Gamma$, $\Sigma$, $\Psi$, and $\Omega$ depend only on
the protocol shape $\mathcal{C}$. The efficiency can be optimized
hierarchically with respect to the protocol shape $\mathcal{C}$ in
parameter space and the cycle duration $\tau$, respectively. To determine
the maximal efficiency for a fixed cycle shape, we optimize $\varepsilon$
with respect to the cycle duration $\tau$. Taking the derivative
yields a physically relevant solution
\begin{equation}
\tau^{*}=\frac{\Sigma}{\Gamma}+\sqrt{\frac{\Sigma^{2}}{\Gamma^{2}}+\frac{\Sigma\Psi}{\Gamma\Omega}}.\label{seq:optimtimewo}
\end{equation}
Substituting it back into the efficiency expression leads to
\begin{equation}
\varepsilon_{\mathrm{max}}=r\frac{\sqrt{\text{\ensuremath{\phi}}+1}-1}{\sqrt{\text{\ensuremath{\phi}}+1}+1},\label{seq:maeffc}
\end{equation}
where $r(\mathcal{C})=\Gamma/\Psi$ characterizes the asymmetry between
geometric work extraction and active energetic cost, whereas $\phi(\mathcal{C})=\Gamma\Psi/(\Sigma\Omega)$
plays the role of an effective figure of merit. Unlike conventional
thermoelectric systems, both quantities depend explicitly on the protocol
shape through the geometric and dissipative contributions of the cycle.
A second optimization stage is therefore required, in which the protocol
shape is varied to maximize $\varepsilon_{\mathrm{max}}(\mathcal{C})$.
Since no analytical solution is available for this shape optimization,
the optimal efficiency protocol is determined numerically. In the
calculations presented in the main text, the protocol shape is parameterized
by a finite set of control variables and optimized using a genetic
algorithm \citep{Goldberg2012}.

\subsection{Optimization of power\label{SSec-fiveC}}

The output power of the active machine is
\begin{equation}
P(\mathcal{C},\tau)\equiv\frac{W(\mathcal{C},\tau)}{\tau}=\frac{\Gamma(\mathcal{C})}{\tau}-\frac{\Sigma(\mathcal{C})}{\tau^{2}}.\label{eq:protpow}
\end{equation}
For a fixed protocol shape, maximizing the power with respect to the
cycle duration yields
\begin{equation}
\tau^{*}=\frac{2\Sigma}{\Gamma}.\label{seq:oppotidu}
\end{equation}
Substituting this result back into the power expression gives
\begin{equation}
P_{\mathrm{max}}=\frac{\Gamma^{2}}{4\Sigma}.\label{seq:maoppow}
\end{equation}
The quantity $\Gamma$ measures the geometric work generated by the
thermodynamic curvature $\mathcal{F}$ enclosed by the cycle, whereas
$\Sigma$ quantifies the dissipation associated with the thermodynamic
metric $g$. Maximizing the power therefore amounts to identifying
protocol shapes that optimally balance curvature-induced work extraction
against metric-induced dissipation. In the numerical calculations
presented in the main text, the protocol shape is parameterized by
a finite set of variables and optimized using a genetic algorithm.
The resulting optimal cycles correspond to maxima of the functional
$\Gamma^{2}/\Sigma$.

\section{Active Brownian particle under harmonic confinement\label{SSec-four}}

In this section, we consider a two-dimensional active Brownian particle,
whose position vector is denoted by $\mathbf{r}=(x,y)$, confined
by the harmonic potential
\begin{equation}
U(x,y,\boldsymbol{\lambda})=\frac{\lambda_{1}}{2}(x^{2}+y^{2})+\lambda_{2}xy,\label{seq:harmoncon}
\end{equation}
where $\boldsymbol{\lambda}(t)\equiv(\lambda_{1},\lambda_{2})$ represents
the set of externally controlled parameters. Stability of the confinement
requires $\lambda_{1}>|\lambda_{2}|$. The dynamics of the particle
is governed by the Langevin equations
\begin{align}
\dot{x} & =-\lambda_{1}x-\lambda_{2}y+v\cos\theta+\xi_{x}(t),\nonumber \\
\dot{y} & =-\lambda_{2}x-\lambda_{1}y+v\sin\theta+\xi_{y}(t).\label{seq:langevinequhar}
\end{align}
Throughout this example, we set the friction coefficient $\gamma=1$
for simplicity.

\subsection{Solution of the Langevin equation\label{SSec-fourA}}

To evaluate the correlation functions entering the response kernels,
we first solve the Langevin equations$\ $(\ref{seq:langevinequhar}).
Introducing the vector notation $\mathbf{e}=(\cos\theta\text{\ }\sin\theta)^{T}$,
the dynamics can be written compactly as
\begin{equation}
\dot{\mathbf{r}}=-A\mathbf{r}+v\mathbf{e}+\boldsymbol{\xi},\label{seq:vectorlangevin}
\end{equation}
where
\begin{equation}
A=\left(\begin{array}{cc}
\lambda_{1} & \lambda_{2}\\
\lambda_{2} & \lambda_{1}
\end{array}\right).\label{seq:harstrength}
\end{equation}
For fixed control parameters $\boldsymbol{\lambda}$, the formal solution
is
\begin{equation}
\mathbf{r}(t)=\mathbf{r}(0)e^{-A(t-t_{0})}+\int_{t_{0}}^{t}dse^{-A(t-s)}(v\mathbf{e}(s)+\boldsymbol{\xi}(s)).\label{seq:formalsoultion}
\end{equation}
Taking the stationary limit $t_{0}\to-\infty$, the contribution from
the initial condition decays exponentially and vanishes, yielding
\begin{equation}
\mathbf{r}(t)=\int_{t_{0}}^{t}dse^{-A(t-s)}(v\mathbf{e}(s)+\boldsymbol{\xi}(s)).\label{seq:stationlimit}
\end{equation}
The matrix $A$ is diagonalizable with eigenvalues $\alpha_{\pm}=\lambda_{1}\pm|\lambda_{2}|$,
and can be decomposed as
\begin{equation}
A=P\left(\begin{array}{cc}
\alpha_{+} & 0\\
0 & \alpha_{-}
\end{array}\right)P^{T},\label{seq:Adecomn}
\end{equation}
where
\begin{equation}
P=\frac{1}{\sqrt{2}}\left(\begin{array}{cc}
1 & 1\\
1 & -1
\end{array}\right).\label{seq:eigmatrix}
\end{equation}
Using the spectral mapping theorem, the matrix exponential becomes
\begin{equation}
e^{-At}=P\left(\begin{array}{cc}
e^{-\alpha_{+}t} & 0\\
0 & e^{-\alpha_{-}t}
\end{array}\right)P^{T}=\frac{1}{2}\left(\begin{array}{cc}
e^{-\alpha_{+}t}+e^{-\alpha_{-}t} & e^{-\alpha_{+}t}-e^{-\alpha_{-}t}\\
e^{-\alpha_{+}t}-e^{-\alpha_{-}t} & e^{-\alpha_{+}t}+e^{-\alpha_{-}t}
\end{array}\right),\label{seq:ematrixdecom}
\end{equation}
which can be written explicitly as
\begin{equation}
e^{-At}=\frac{1}{2}\left(\begin{array}{cc}
e^{-\alpha_{+}t}+e^{-\alpha_{-}t} & e^{-\alpha_{+}t}-e^{-\alpha_{-}t}\\
e^{-\alpha_{+}t}-e^{-\alpha_{-}t} & e^{-\alpha_{+}t}+e^{-\alpha_{-}t}
\end{array}\right).\label{seq:eacorrfunex}
\end{equation}

\subsection{Correlation functions\label{SSec-fourB}}

Using the solution obtained above, the correlation functions required
for evaluating the response kernels can be derived analytically.

The coordinate correlation functions are
\begin{align}
C_{xx}(t) & =\langle x(t)x(0)\rangle=\frac{K_{+}}{4(\lambda_{1}+|\lambda_{2}|)}e^{-\alpha_{+}t}+\frac{K_{-}}{4(\lambda_{1}-|\lambda_{2}|)}e^{-\alpha_{-}t},\nonumber \\
C_{xy}(t) & =\langle x(t)y(0)\rangle=\frac{K_{+}}{4(\lambda_{1}+|\lambda_{2}|)}e^{-\alpha_{+}t},\nonumber \\
C_{yx}(t) & =\langle y(t)x(0)\rangle=\frac{K_{+}}{4(\lambda_{1}+|\lambda_{2}|)}e^{-\alpha_{+}t},\nonumber \\
C_{yy}(t) & =\langle y(t)y(0)\rangle=\frac{K_{+}}{4(\lambda_{1}+|\lambda_{2}|)}e^{-\alpha_{+}t}+\frac{K_{-}}{4(\lambda_{1}-|\lambda_{2}|)}e^{-\alpha_{-}t},\label{seq:coordinatecorr}
\end{align}
where $K_{\pm}\equiv2T+v^{2}\tau_{p}/(1-\tau_{p}^{2}\alpha_{\pm}^{2})$.
The correlations between the particle coordinates and the self-propulsion
velocity are
\begin{align}
D_{xx}(t) & =v\langle x(t)e_{x}(0)\rangle=\frac{R_{+}}{2}e^{-\alpha_{+}t}+\frac{R_{-}}{2}e^{-\alpha_{-}t},\nonumber \\
D_{xy}(t) & =v\langle x(t)e_{y}(0)\rangle=\frac{R_{+}}{2}e^{-\alpha_{+}t}-\frac{R_{-}}{2}e^{-\alpha_{-}t},\nonumber \\
D_{yx}(t) & =v\langle y(t)e_{x}(0)\rangle=\frac{R_{+}}{2}e^{-\alpha_{+}t}-\frac{R_{-}}{2}e^{-\alpha_{-}t},\nonumber \\
D_{yy}(t) & =v\langle y(t)e_{y}(0)\rangle=\frac{R_{+}}{2}e^{-\alpha_{+}t}+\frac{R_{-}}{2}e^{-\alpha_{-}t},\label{seq:xvcorr}
\end{align}
with $R_{\pm}\equiv v^{2}\tau_{p}/(1-\tau_{p}^{2}\alpha_{\pm}^{2})$.
The correlations between the self-propulsion velocity and the particle
coordinates are
\begin{align}
E_{xx}(t) & =v\langle e_{x}(t)x(0)\rangle=\frac{v^{2}}{4}(\frac{1}{\alpha_{+}}+\frac{1}{\alpha_{-}})e^{-t/\tau_{p}},\nonumber \\
E_{xy}(t) & =v^{2}\langle e_{x}(t)y(0)\rangle=\frac{v^{2}}{4}(\frac{1}{\alpha_{+}}-\frac{1}{\alpha_{-}})e^{-t/\tau_{p}},\nonumber \\
E_{yx}(t) & =v^{2}\langle e_{y}(t)x(0)\rangle=\frac{v^{2}}{4}(\frac{1}{\alpha_{+}}-\frac{1}{\alpha_{-}})e^{-t/\tau_{p}},\nonumber \\
E_{yy}(t) & =v^{2}\langle e_{y}(t)y(0)\rangle=\frac{v^{2}}{4}(\frac{1}{\alpha_{+}}+\frac{1}{\alpha_{-}})e^{-t/\tau_{p}}.\label{seq:vxcorre}
\end{align}
Finally, the correlations between the thermal noise and the particle
coordinates are
\begin{align}
F_{xx}(t) & =\langle\xi_{x}(t)x(0)\rangle=T(\frac{1}{\alpha_{+}}+\frac{1}{\alpha_{-}})\delta(t),\nonumber \\
F_{xy}(t) & =\langle\xi_{x}(t)y(0)\rangle=T(\frac{1}{\alpha_{+}}-\frac{1}{\alpha_{-}})\delta(t),\nonumber \\
F_{yx}(t) & =\langle\xi_{y}(t)x(0)\rangle=T(\frac{1}{\alpha_{+}}-\frac{1}{\alpha_{-}})\delta(t),\nonumber \\
F_{yy}(t) & =\langle\xi_{y}(t)y(0)\rangle=T(\frac{1}{\alpha_{+}}+\frac{1}{\alpha_{-}})\delta(t).\label{seq:thenoxcorr}
\end{align}

\subsection{Response kernels\label{SSec-fourC}}

Substituting the correlation functions derived in the previous subsection
into Eq.$\ $(\ref{seq:resknerl}), the response kernels associated
with the generalized force $\partial U/\partial\boldsymbol{\lambda}$
can be expressed as
\begin{align}
I_{1}(\partial U/\partial\lambda_{1})= & \frac{1}{2T}\int_{0}^{\infty}dt[(1+2\lambda_{1}t)(C_{xx}^{2}(t)+C_{xy}^{2}(t)+C_{yx}^{2}(t)+C_{yy}^{2}(t))+2\lambda_{2}t(C_{xx}(t)C_{xy}(t)+C_{yx}(t)C_{yy}(t))\nonumber \\
 & -t(C_{xx}(t)D_{xx}(t)+C_{xy}(t)D_{xy}(t)+C_{yx}(t)D_{yx}(t)+C_{yy}(t)D_{yy}(t))],\nonumber \\
I_{1}(\partial U/\partial\lambda_{2})= & \frac{1}{2T}\int_{0}^{\infty}dt[(1+2\lambda_{1}t)(C_{xx}(t)C_{xy}(t)+C_{yx}(t)C_{yy}(t))+\lambda_{2}t(C_{xx}^{2}(t)+C_{xy}^{2}(t)+C_{yx}^{2}(t)+C_{yy}^{2}(t))\nonumber \\
 & -t(C_{xy}(t)D_{xx}(t)+C_{xx}(t)D_{xy}(t)+C_{yy}(t)D_{yx}(t)+C_{yx}(t)D_{yy}(t))],\nonumber \\
I_{2}(\partial U/\partial\lambda_{1})= & \frac{1}{2T}\int_{0}^{\infty}dt[(1+2\lambda_{1}t)(C_{xx}(t)C_{yx}(t)+C_{xy}(t)C_{yy}(t))+2\lambda_{2}t(C_{xx}(t)C_{yy}(t)+C_{xy}(t)C_{yx}(t))\nonumber \\
 & -t(C_{xx}(t)D_{yx}(t)+C_{yx}(t)D_{xx}(t)+C_{xy}(t)D_{yy}(t)+C_{yy}(t)D_{xy}(t))],\nonumber \\
I_{2}(\partial U/\partial\lambda_{2})= & \frac{1}{2T}\int_{0}^{\infty}dt[(1+2\lambda_{1}t)(C_{xx}(t)C_{yy}(t)+C_{xy}(t)C_{yx}(t))+2\lambda_{2}t(C_{xx}(t)C_{yx}(t)+C_{xy}(t)C_{yy}(t))\nonumber \\
 & -t(C_{xy}(t)D_{yx}(t)+C_{yy}(t)D_{xx}(t)+C_{xx}(t)D_{yy}(t)+C_{yx}(t)D_{xy}(t))].\label{seq:IUlambdacorre}
\end{align}
Because all correlation functions consist of exponential modes, the
above time integrals can be evaluated analytically. After straightforward
calculations, we obtain
\begin{align}
I_{1}(\partial U/\partial\lambda_{1})= & \frac{1}{2T}\{\frac{K_{+}^{2}[1+(\lambda_{1}+\lambda_{2}/2)/(\lambda_{1}+|\lambda_{2}|)]}{8(\lambda_{1}+|\lambda_{2}|)^{3}}+\frac{K_{-}^{2}[1+\lambda_{1}/(\lambda_{1}-|\lambda_{2}|)]}{16(\lambda_{1}-|\lambda_{2}|)^{3}}+\frac{K_{+}K_{-}[1+\lambda_{2}/(4\lambda_{1})]}{4\lambda_{1}(\lambda_{1}^{2}-\lambda_{2}^{2})}\nonumber \\
 & -\frac{K_{+}R_{+}}{8(\lambda_{1}+|\lambda_{2}|)^{3}}-\frac{K_{-}R_{-}}{16(\lambda_{1}-|\lambda_{2}|)^{3}}-\frac{K_{-}R_{+}}{16\lambda_{1}^{2}(\lambda_{1}-|\lambda_{2}|)}\},\nonumber \\
I_{1}(\partial U/\partial\lambda_{2})= & \frac{1}{2T}\{\frac{K_{+}^{2}[1+(\lambda_{1}+\lambda_{2})/(\lambda_{1}+|\lambda_{2}|)]}{16(\lambda_{1}+|\lambda_{2}|)^{3}}+\frac{K_{+}K_{-}[1+\lambda_{2}/(2\lambda_{1})]}{8\lambda_{1}(\lambda_{1}^{2}-\lambda_{2}^{2})}+\frac{K_{-}^{2}\lambda_{2}}{32(\lambda_{1}-|\lambda_{2}|)^{4}}\nonumber \\
 & -\frac{K_{+}R_{+}}{8(\lambda_{1}+|\lambda_{2}|)^{3}}+\frac{K_{-}R_{-}}{16(\lambda_{1}-|\lambda_{2}|)^{3}}-\frac{K_{-}R_{+}}{16\lambda_{1}^{2}(\lambda_{1}-|\lambda_{2}|)}\},\nonumber \\
I_{2}(\partial U/\partial\lambda_{1})= & \frac{1}{2T}\{\frac{K_{+}^{2}[1+(\lambda_{1}+\lambda_{2})/(\lambda_{1}+|\lambda_{2}|)]}{16(\lambda_{1}+|\lambda_{2}|)^{3}}+\frac{K_{+}K_{-}[1+\lambda_{2}/(2\lambda_{1})]}{8\lambda_{1}(\lambda_{1}^{2}-\lambda_{2}^{2})}+\frac{K_{-}^{2}\lambda_{2}}{32(\lambda_{1}-|\lambda_{2}|)^{4}}\nonumber \\
 & -\frac{K_{+}R_{+}}{8(\lambda_{1}+|\lambda_{2}|)^{3}}+\frac{K_{-}R_{-}}{16(\lambda_{1}-|\lambda_{2}|)^{3}}-\frac{K_{-}R_{+}}{16\lambda_{1}^{2}(\lambda_{1}-|\lambda_{2}|)}\},\nonumber \\
I_{2}(\partial U/\partial\lambda_{2})= & \frac{1}{2T}\{\frac{K_{+}^{2}[1+(\lambda_{1}+\lambda_{2})/(\lambda_{1}+|\lambda_{2}|)]}{16(\lambda_{1}+|\lambda_{2}|)^{3}}+\frac{K_{-}^{2}[1+\lambda_{1}/(\lambda_{1}-|\lambda_{2}|)]}{32(\lambda_{1}-|\lambda_{2}|)^{3}}+\frac{K_{+}K_{-}[1+\lambda_{2}/(2\lambda_{1})]}{8\lambda_{1}(\lambda_{1}^{2}-\lambda_{2}^{2})}\nonumber \\
 & -\frac{K_{+}R_{+}}{8(\lambda_{1}+|\lambda_{2}|)^{3}}-\frac{K_{-}R_{-}}{16(\lambda_{1}-|\lambda_{2}|)^{3}}-\frac{K_{-}R_{+}}{16\lambda_{1}^{2}(\lambda_{1}-|\lambda_{2}|)}\}.\label{seq:IUlambdaan}
\end{align}
These expressions completely determine the dissipation metric and
yield the rescaled dissipation $\Sigma\equiv\int_{0}^{1}[\boldsymbol{\lambda}']^{T}\cdot\boldsymbol{I}(\partial U/\partial\boldsymbol{\lambda})\cdot[\boldsymbol{\lambda}']du$.
Applying the same procedure to the active-force contribution $f$,
we obtain the response kernels $\boldsymbol{I}(f)$ as
\begin{align}
I_{1}(f)= & \frac{1}{2T}\{-\frac{v^{2}}{2}[\frac{K_{+}}{2(\lambda_{1}+|\lambda_{2}|)(\alpha_{+}+1/\tau_{p})}+\frac{\lambda_{1}^{2}+\lambda_{2}|\lambda_{2}|}{\lambda_{1}+|\lambda_{2}|}\frac{\lambda_{1}K_{-}}{2(\lambda_{1}-|\lambda_{2}|)^{2}(\alpha_{-}+1/\tau_{p})}]\nonumber \\
 & -\frac{v^{2}}{2}[\frac{(\lambda_{1}+\lambda_{2})^{2}K_{+}}{(\lambda_{1}+|\lambda_{2}|)^{2}(\alpha_{+}+1/\tau_{p})^{2}}+\frac{\lambda_{1}}{\lambda_{1}+|\lambda_{2}|}\frac{(\lambda_{1}^{2}+\lambda_{2}^{2}-2\lambda_{2}|\lambda_{2}|)K_{-}}{(\lambda_{1}-|\lambda_{2}|)^{2}(\alpha_{-}+1/\tau_{p})^{2}}]\nonumber \\
 & +\frac{\lambda_{1}^{2}+\lambda_{2}^{2}}{(\lambda_{1}^{2}-\lambda_{2}^{2})^{2}}\frac{v^{4}\tau_{p}}{4}(1+\lambda_{1}\tau_{p})-\frac{\lambda_{1}|\lambda_{2}|}{(\lambda_{1}^{2}-\lambda_{2}^{2})^{2}}\frac{v^{4}\tau_{p}^{2}\lambda_{2}}{2}+\frac{\lambda_{1}^{2}+\lambda_{2}^{2}}{(\lambda_{1}^{2}-\lambda_{2}^{2})^{2}}2v^{2}D\nonumber \\
 & +\frac{v^{2}}{2}[\frac{(\lambda_{1}+\lambda_{2})(K_{+}+2R_{+})}{2(\lambda_{1}+|\lambda_{2}|)(\alpha_{+}+1/\tau_{p})^{2}}+\frac{\lambda_{1}K_{-}+2(\lambda_{1}-\lambda_{2})R_{-}}{2(\lambda_{1}-|\lambda_{2}|)(\alpha_{-}+1/\tau_{p})^{2}}]-\frac{\lambda_{1}}{\lambda_{1}^{2}-\lambda_{2}^{2}}\frac{v^{4}\tau_{p}^{2}}{4}\},\nonumber \\
I_{2}(f)= & \frac{1}{2T}\{-\frac{v^{2}}{2}[\frac{(\lambda_{1}+\lambda_{2})K_{+}}{2(\lambda_{1}+|\lambda_{2}|)^{2}(\alpha_{+}+1/\tau_{p})}-\frac{\lambda_{1}}{\lambda_{1}+|\lambda_{2}|}\frac{(|\lambda_{2}|-\lambda_{2})K_{-}}{2(\lambda_{1}-|\lambda_{2}|)^{2}(\alpha_{-}+1/\tau_{p})}]\nonumber \\
 & -\frac{v^{2}}{2}[\frac{(\lambda_{1}+\lambda_{2})^{2}K_{+}}{(\lambda_{1}+|\lambda_{2}|)^{2}(\alpha_{+}+1/\tau_{p})^{2}}-\frac{1}{\lambda_{1}+|\lambda_{2}|}\frac{(\lambda_{1}^{2}|\lambda_{2}|+\lambda_{2}^{2}|\lambda_{2}|-2\lambda_{1}^{2}\lambda_{2})K_{-}}{(\lambda_{1}-|\lambda_{2}|)^{2}(\alpha_{-}+1/\tau_{p})^{2}}]\nonumber \\
 & -\frac{\lambda_{1}|\lambda_{2}|}{(\lambda_{1}^{2}-\lambda_{2}^{2})^{2}}\frac{v^{4}\tau_{p}}{2}(1+\lambda_{1}\tau_{p})+\frac{\lambda_{1}^{2}+\lambda_{2}^{2}}{(\lambda_{1}^{2}-\lambda_{2}^{2})^{2}}\frac{v^{4}\tau_{p}^{2}\lambda_{2}}{4}-\frac{2\lambda_{1}|\lambda_{2}|}{(\lambda_{1}^{2}-\lambda_{2}^{2})^{2}}2v^{2}D\nonumber \\
 & +\frac{v^{2}}{2}[\frac{(\lambda_{1}+\lambda_{2})K_{+}+4\lambda_{1}R_{+}}{2(\lambda_{1}+|\lambda_{2}|)(\alpha_{+}+1/\tau_{p})^{2}}+\frac{\lambda_{2}K_{-}}{2(\lambda_{1}-|\lambda_{2}|)(\alpha_{-}+1/\tau_{p})^{2}}]-\frac{|\lambda_{2}|}{\lambda_{1}^{2}-\lambda_{2}^{2}}\frac{v^{4}\tau_{p}^{2}}{4}\},\label{seq:Ifreskern}
\end{align}
which determine the excess active energy input generated by a finite-time
protocol, $\Psi\equiv\int_{0}^{1}\boldsymbol{I}(f)\cdot[\boldsymbol{\lambda}']du$.

The quasistatic output work can likewise be evaluated analytically.
Using Stokes' theorem, it can be expressed as $\Gamma=-\sum_{\mu}\oint_{\mathcal{C}}\langle\partial_{\mu}U\rangle_{s}d\lambda_{\mu}=\sum_{\mu\nu}\iint\mathcal{F}_{\mu\nu}d\lambda_{\mu}\wedge d\lambda_{\nu}$,
where the only nonvanishing component of the thermodynamic curvature
is
\begin{equation}
\mathcal{F}_{12}=\frac{K_{+}[1-\mathrm{sgn}(\lambda_{2})]}{4(\lambda_{1}+|\lambda_{2}|)^{2}}-\frac{\tau_{p}}{2v^{2}}K_{+}R_{+}[1-\mathrm{sgn}(\lambda_{2})]+\frac{K_{-}\mathrm{sgn}(\lambda_{2})}{4(\lambda_{1}-|\lambda_{2}|)^{2}}-\frac{\tau_{p}}{2v^{2}}K_{-}R_{-}\mathrm{sgn}(\lambda_{2}).\label{seq:fcurveture}
\end{equation}
Finally, the steady-state energy injection rate due to self-propulsion
is given by $\Omega\equiv\int_{0}^{1}\langle f\rangle_{s}du$ with
\begin{align}
\langle f\rangle_{s} & =-\lambda_{1}(R_{+}+R_{-})-\lambda_{2}(R_{+}-R_{-})+v^{2}.\label{seq:averf}
\end{align}
Together with the dissipation metric and thermodynamic curvature derived
above, these analytical results completely characterize the geometric
thermodynamics of the harmonically confined active Brownian particle.

\subsection{Numerical optimization of the output work, power, and efficiency\label{SSec-fourD}}

The analytical expressions derived above completely determine the
quasistatic work $\Gamma(\mathcal{C})$, the rescaled dissipation
$\Sigma(\mathcal{C})$, the excess active energetic cost $\Psi(\mathcal{C})$,
and the steady-state active energy input $\Omega(\mathcal{C})$. Consequently,
the performance of the active machine can be evaluated for any prescribed
protocol shape. To determine the optimal cycles shown in the main
text, the protocol shape was parameterized by a finite set of variables.
Specifically, the control protocol was represented as
\begin{equation}
\lambda_{\mu}(u)=a_{\mu}^{(0)}+\sum_{n=1}^{N_{F}}[a_{\mu}^{(n)}\cos(2\pi nu)+b_{\mu}^{(n)}\sin(2\pi nu)],\label{seq:parapro2}
\end{equation}
where $u=t/\tau\in[0,1]$ is the dimensionless cycle parameter and
$(a_{\mu}^{(n)},b_{\mu}^{(n)})$ constitute the optimization variables.
For a given set of parameters, the geometric quantities $\Gamma$,
$\Sigma$, $\Psi$, and $\Omega$ were evaluated numerically using
the analytical expressions derived above. The objective functions
were chosen as
\begin{equation}
J_{W}=\Gamma-\frac{\Sigma}{\tau}\label{seq:opworob}
\end{equation}
for work optimization,

\begin{equation}
J_{\varepsilon}=r\frac{\sqrt{1+\phi}-1}{\sqrt{1+\phi}+1}\label{seq:maxeffob}
\end{equation}
for efficiency optimization, and

\begin{equation}
J_{P}=\frac{\Gamma^{2}}{4\Sigma}\label{seq:maxpoob}
\end{equation}
for power optimization. The resulting optimization problems were solved
using a genetic algorithm. The optimization was performed subject
to the stability condition $\lambda_{1}(u)>|\lambda_{2}(u)|$ throughout
the cycle. The optimal protocols reported in the main text correspond
to the converged solutions obtained after repeated runs with different
initial populations.

The resulting optimal cycles reveal how thermodynamic curvature and
dissipation compete in determining machine performance. Protocols
maximizing work tend to enclose regions of large curvature, whereas
protocols maximizing power or efficiency shift toward regions where
the gain from geometric work extraction is balanced against the dissipation
encoded by the thermodynamic metric.

\bibliographystyle{apsrev4-1}
\bibliography{ref}